\documentclass[preprint2]{aastex}
\usepackage{graphicx}

\slugcomment{To be submitted to \emph{Astrophysical Journal}}

\begin{document}

\title{Synchrotron emission driven by the Cherenkov-drift instability in active galactic nuclei.}

\author{Z. Osmanov}
\affil{Free University of Tbilisi, 0183, Tbilisi, Georgia}
\email{z.osmanov@freeuni.edu.ge }

\and

\author{N. Chkheidze }
\affil{Center for theoretical Astrophysics, ITP, Ilia State
University, 0162, Tbilisi, Georgia}

\begin{abstract}
In the present paper we study generation of the synchrotron emission
by means of the feedback of Cherenkov drift waves  on the particle
distribution via the diffusion process. It is shown that despite the
efficient synchrotron losses the excited Cherenkov drift instability
leads to the quasi-linear diffusion (QLD), effect of which is
balanced by dissipation factors and as a result the pitch angles are
prevented from damping, maintaining the corresponding synchrotron
emission. The model is analyzed for a wide range of physical
parameters and it is shown that the mechanism of QLD guarantees the
generation of electromagnetic radiation from soft $X$-rays up to
soft $\gamma$-rays, strongly correlated with Cherenkov drift
emission ranging from IR up to UV energy domains.
\end{abstract}


\section{Introduction }

Recently an interest to $X$-ray and $\gamma$-ray emission from
active galactic nuclei (AGNs) has substantially increased due to new
observational data from satellite telescopes. Investigations of
cosmic objects passed to a new stage when in $2007$ the AGILE
(Astro-rivelatore Gamma a Immagini Leggero) satellite and in $2008$
the Fermi spacecraft were launched. These new observational results
are of fundamental importance in studying the emission properties of
several $X$-ray and $\gamma$-ray sources, like AGNs, pulsars, gamma
ray bursts etc.

Most commonly it is believed that highly relativistic electrons do
not participate in the synchrotron emission. Energy losses are so
efficient that particles very rapidly loose their energies and
transit to the ground Landau states. In spite of this, as it was
shown in a series of articles by
\cite{difus6,difus4,difus5,cheren1,difus3}, the synchrotron
radiation still might account for the radiation processes produced
by very energetic particles in AGNs.


According to \cite{kmm} in strong magnetic fields the plasma may
induce the unstable cyclotron waves. On the other hand, as it was
shown by \cite{lomin}, these waves via the QLD affect the particle
distribution as along as across the magnetic field lines. Such a
feedback of cyclotron waves on particles will inevitably lead to the
creation of the pitch angles, restoring the synchrotron emission.
Apart from AGNs the process of the QLD was applied to pulsars as
well and it was shown that the mentioned mechanism is very efficient
for pulsar magnetospheres
\citep{lomin,machus1,malmach,difus,difus1}.

This approach makes possible for the synchrotron radiation to be a
working mechanism despite strong energy losses. A very interesting
property of the QLD is that it enables to produce highly correlated
radiation in two different energy bands. In this process one has two
major forces that influence the particle distribution function. On
the one hand, the diffusion attempts to increase the values of the
pitch angles of resonant particles and, and on the other hand, the
emitting particles are affected by dissipative forces intending to
decrease their pitch angles. It was shown in the papers listed above
that under certain conditions the dissipative and diffusion factors
balance each other, the physical system reaches stationary state and
the pitch angles saturate. Consequently,the cyclotron emission by
means of the diffusion redistributes resonant particles and as a
result, the synchrotron emission is produced. This mechanism
guarantees strongly correlated radiation in two different energy
bands.

In the framework of this emission model, the radiation is generated
in two energy bands because of the feedback of the excited cyclotron
waves on particles by means of the QLD, which start to radiate in
the synchrotron regime. In general similar emission mechanism can be
driven not only via the cyclotron waves. Particularly, in the
present paper, unlike the aforementioned articles, we consider the
feedback of the Cherenkov-drift waves on the resonant particles via
the QLD. This process should also switch the synchrotron radiation
mechanism and must provide correlated emission in different energy
domains.

The paper is arranged in the following way. In section~II, we
introduce the theory of the quasi-linear diffusion. In section~III,
we apply the model to AGNs and in section~IV, we summarize our
results.

\section[]{Model}

In this section we develop the model of the QLD of particles driven
by means of the feedback of the Cherenkov drift modes in the
magnetospheres of AGNs. Normally, in the mentioned region, with the
typical lengthscales $l\sim 10^{13-14}cm$, values of the Lorentz
factors of electrons may vary from $\sim 1$ to $\sim 10^7$. This
range for the Lorentz factors was implied in \cite{osm7,ra08}, where
where it was shown that acceleration of particles becomes extremely
efficient in the light cylinder zone (a hypothetical area, where the
linear velocity of rotation exactly equals the speed of light) due
to the relativistic effects of rotation. We apply methods developed
in the mentioned papers in the framework of the present article. For
simplicity one can assume that the magnetosphere of AGN is composed
of a low energy plasma component with the Lorentz factors
$\gamma_p$, and a highly relativistic part - the beam component with
$\gamma_b$. As we have already mentioned, the synchrotron radiation
is suppressed for strongly relativistic electrons. In particular,
one can see that the cooling timescale of beam electrons is given by
$t_{syn}\sim\gamma_b mc^2/P_{syn}$, where $m$ is the electron's
mass, $c$ is the speed of light, $P_{syn}\approx
2e^4\gamma_b^2B^2/3m^2c^3$ is the single particle synchrotron
emission power, $B$ is the magnetic induction and $e$ is the
electron's charge. By taking into account typical values of the
magnetic field in an ambient close to the supermassive black hole
(SMBH), $B\approx 10^{1-4}$G, one can see that the cooling timescale
varies in the following interval $5\times (10^{-7}-1)$s. On the
other hand, close to the light cylinder zone, due to the efficient
curvature drift instability, the magnetic field lines significantly
twist \citep{cdiagn}, therefore, the lengthscale is still of the
order of $l$ and the corresponding plasma escape timescale, $l/c$,
is of the order of $3\times 10^{2-3}$s. As we see, the synchrotron
cooling timescale is much less than the kinematic timescale,
therefore, in AGN magnetospheres, due to the strong energy losses,
particles very rapidly transit to the ground Landau states resulting
in the termination of the emission process.

Generally speaking, as was explained by \cite{shapo1}, the necessary
condition for the development of the Cherenkov drift instability
(ChDI) is the presence of the beam component in plasmas. Since the
magnetic field lines are always curved, it is evident that the
electrons will drift along a direction perpendicular to the plane of
the curved magnetic field lines with the following velocity
\begin{equation}\label{ud}
u_x = \frac{\gamma_bc^2}{\omega_B\rho},
\end{equation}
where $\omega_B = eB/(mc)$ is the cyclotron frequency of the
particle and $\rho$ is the curvature radius of the magnetic field
line. It is clear that for highly relativistic particles
($\gamma_b>>1$) the drift velocity becomes significant and the ChDI
with the following resonance condition arises \citep{shapo1}
\begin{equation}\label{res}
\omega-k_{_{\parallel}}\upsilon_{_{\parallel}}-k_xu_x = 0,
\end{equation}
where $k_{_{\parallel}}$ and $\upsilon_{_{\parallel}}$ are the
longitudinal components (along the magnetic field line) respectively
of the wave vector and the particle velocity, and $k_x$ is the wave
vector's component along the drift direction.

During the process of ChDI the transverse (t) waves might be
excited, having the growth rate \citep{shapo1}
\begin{equation}\label{incr}
\Gamma_k =
\frac{\pi}{2}\frac{\omega_b^2}{\omega}\frac{\gamma_b}{\gamma_p^2}A_k,
\end{equation}
where
\begin{equation}\label{om}
\omega = \frac{\omega_b\gamma_bc}{\gamma_p^{3/2}u_x}
\end{equation}
is the frequency of Cherenkov emission, $\omega_b = \sqrt{4\pi
n_be^2/m}$ is the Langmuir frequency, $n_b$ is the beam number
density, $A_k = (k_r/k_x)^2$ and $k_r$ is the component of the wave
vector along the curvature radius of magnetic field lines.
Throughout the paper we assume $A_k = 1$ \citep{kmm}.

During their motion in a nonuniform magnetic field the charged
relativistic particles undergo two major dissipative forces: $\bf H$
- responsible for conservation of the adiabatic invariant
\citep{landau}
\begin{equation}\label{hper}
H_{\perp} = -\frac{c}{\rho}p_{\perp},\;\;\;\; H_{\parallel} =
\frac{c}{\rho p_{\parallel}}p_{\perp}^2,
\end{equation}
and the synchrotron radiation reaction force \citep{landau}
\begin{equation}\label{f}
    F_{\perp}=-\alpha\frac{p_{\perp}}{p_{\parallel}}\left(1+\frac{p_{\perp}^{2}}{m^{2}c^{2}}\right),
    F_{\parallel}=-\frac{\alpha}{m^{2}c^{2}}p_{\perp}^{2},
\end{equation}
where $\alpha=2e^{2}\omega_{B}^{2}/3c^{2}$, $p_{\perp}$ and
$p_{\parallel}$ are transversal and longitudinal components of the
momentum, respectively.

The effect of the aforementioned dissipative forces decreases the
pitch angles, which inevitably causes attenuation of the
corresponding synchrotron emission process. The situation
drastically changes due to diffusion as along as across the magnetic
field lines, since this process tries to increase the pitch angles.
Under certain conditions diffusion and dissipation forces might
balance each other and as a result the synchrotron radiation process
will be maintained. The QLD influences the particle distribution
function and the corresponding kinetic equation can be written as
\citep{ninoz,machus1,malmach}
\begin{eqnarray}\label{kin1}
\frac{\partial\textit{f }^{0}\left(\mathbf{p}\right)}{\partial
    t}+\frac{\partial}{\partial
p_{\parallel}}\left\{\left[
    F_{\parallel}+H_{\parallel}\right]\textit{f
}^{0}\left(\mathbf{p}\right)\right\}+\nonumber
\\+\frac{1}{p_{\perp}}\frac{\partial}{\partial
p_{\perp}}\left\{p_{\perp}\left[
    F_{\perp}+H_{\perp}\right]\textit{f
}^{0}\left(\mathbf{p}\right)\right\}=\nonumber
\\=\frac{1}{p_{\perp}}\frac{\partial}{\partial p_{\perp}}\left\{p_{\perp}
D_{\perp,\perp}\frac{\partial\textit{f
}^{0}\left(\mathbf{p}\right)}{\partial p_{\perp}}\right\}+\nonumber
\\
+\frac{\partial}{\partial
p_{\parallel}}\left\{D_{\parallel,\parallel}\frac{\partial\textit{f
}^{0}\left(\mathbf{p}\right)}{\partial p_{\parallel}}\right\},
\end{eqnarray}
where $\textit{f }^{0}\left(\mathbf{p}\right)$ is the distribution
function,
\begin{eqnarray}\label{dkoef}
D_{\perp,\perp} =
8\pi\left(\frac{u_x}{c}\right)^8\left(\frac{e}{mc}\right)^2\frac{W}{\Gamma_k},\nonumber
\\D_{\parallel,\parallel} =
8\pi\left(\frac{u_x}{c}\right)^2\left(\frac{e}{mc}\right)^2\frac{W}{\Gamma_k},
\end{eqnarray}
are the diffusion coefficients. Here $W$ is the energy density of
the excited waves and we have taken into account that
$D_{\perp,\parallel}=D_{\parallel,\perp}=0$ \citep{shapo2}.

It should be noted that pitch angles $\psi$ acquired by particles
during the QLD are small enough to apply the condition,
$\partial/\partial p_{\perp}>>\partial/\partial p_{\parallel}$,
which reduces the kinetic equation (\ref{kin1}) to the following
form
\begin{eqnarray} \label{kin2}
    \frac{\partial\textit{f }^{0}}{\partial
    t}+\frac{1}{p_{\perp}}\frac{\partial}{\partial
    p_{\perp}}\left(p_{\perp}\left[
    F_{\perp}+H_{\perp}\right]\textit{f }^{0}\right)
    =\nonumber
    \\=\frac{1}{p_{\perp}}\frac{\partial}{\partial p_{\perp}}\left(p_{\perp}
D_{\perp,\perp}\frac{\partial\textit{f }^{0}}{\partial
p_{\perp}}\right).
\end{eqnarray}

On the light cylinder lengthscales one can simplify Eq. (\ref{kin2})
by estimating the ratio $F_{\perp}/H_{\perp}$. Taking into account
parameters typical for the light cylinder area, one obtains
\begin{eqnarray} \label{ratio}
    \frac{H_{\perp}}{F_{\perp}}\approx 350\times \frac{M_8}{A_{0.1}L_{40}}\times
    \left(\frac{10^{-3}rad}{\psi}\right)^2
    \times\frac{10^7}{\gamma_{b}}\times\frac{R_{lc}}{\rho},
\end{eqnarray}
where $A_{0.1}=\Omega/(0.1\Omega_{max})$, $\Omega$ and
$\Omega_{max}\approx c^3/(GM_{BH})$ are the dimensionless angular
velocity, the actual angular velocity and the maximum angular
velocity of the supermassive black hole, $M_{BH}$ is its mass,
$G\approx6.67\times 10^{-8}$dyne cm$^2$/g$^2$ is the gravitational
constant, $L_{40} = L/10^{40}$erg/s and $L$ are the dimensionless
and actual values of luminosity of AGN, $M_8=M_{BH}/(10^8M_{\odot})$
is its dimensionless mass, $M_{\odot}$ is the solar mass and
$R_{lc}=c/\Omega$ is the light cylinder radius.

In general, it is evident from Eq. (\ref{ratio}) that depending on
physical parameters one has two major regimes:(I) $F_{\perp}\ll
H_{\perp}$ (II) $H_{\perp}\ll F_{\perp}$.

In the framework of the first approximation one can neglect its
contribution of $F_{\perp}$ in Eq. (\ref{kin2}), which for the
stationary case ($\partial/\partial t = 0$) has the following
solution
\begin{equation}\label{ff1}
    \textit{f}_I(p_{\perp})=C exp\left(\int \frac{H_{\perp}}
    {D_{\perp,\perp}}dp_{\perp}\right)=Ce^{-\left(\frac{p_{\perp}}{p_{\perp_{0}}^{I}}
    \right)^{2}},
\end{equation}
where
\begin{equation}\label{pp01}
     p_{\perp_{0}}{_I}=\left(\frac{2\rho D_{\perp,\perp}}{c}\right)^{1/2}.
\end{equation}

As we see from Eq. (\ref{ff1}), particles are distributed
differently for different values of transverse momentum, therefore,
it is natural to examine the average value of $p_{\perp}$ which in
turn defines the mean value of the pitch angles. A straightforward
calculation leads to the following expression
\begin{equation}\label{avp1}
\langle p_{\perp}\rangle_{I}
 = \frac{\int_{0}^{\infty}p_{\perp} \textit{f}_I(p_{\perp})dp_{\perp}}
 {\int_{0}^{\infty}\textit{f}_I(p_{\perp})dp_{\perp}}
= \frac{p_{\perp_{0}}^{I}}{\sqrt{\pi}},
\end{equation}
and the corresponding average value of the pitch angle
\begin{equation}\label{pitch1}
\langle \psi\rangle_{I}
 = \frac{\langle p_{\perp}\rangle_{I}}{p_{\parallel}}= \frac{1}{\sqrt{\pi}}
 \frac{p_{\perp_{0}}^{I}}{p_{\parallel}}.
\end{equation}
For the second case ($H_{\perp}\ll F_{\perp}$), the distribution
function reduces to
\begin{equation}\label{ff2}
    \textit{f}_{II}(p_{\perp})=C exp\left(\int \frac{F_{\perp}}
    {D_{\perp,\perp}}dp_{\perp}\right)=Ce^{-\left(\frac{p_{\perp}}{p^{II}_{\perp_{0}}}
    \right)^{4}},
\end{equation}
where
\begin{equation}\label{pp02}
     p^{II}_{\perp_{0}}=\left(\frac{4\gamma_bm^3c^3D_{\perp,\perp}}{\alpha}\right)^{1/4},
\end{equation}
and the corresponding mean values of the transverse momentum and the
pitch angle respectively are given by
\begin{equation}\label{avp2}
\langle p_{\perp}\rangle_{II}
 = \frac{\int_{0}^{\infty}p_{\perp} \textit{f}_{II}(p_{\perp})dp_{\perp}}
 {\int_{0}^{\infty}\textit{f}_{II}(p_{\perp})dp_{\perp}}
= \frac{\sqrt{\pi}}{4\Gamma (\frac{5}{4})}p^{II}_{\perp_{0}},
\end{equation}
\begin{equation}\label{pitch2}
\langle \psi\rangle_{II}
 = \frac{\langle p_{\perp}\rangle_{II}}{p_{\parallel}}= \frac{\sqrt{\pi}}{4\Gamma (\frac{5}{4})}
 \frac{p^{II}_{\perp_{0}}}{p_{\parallel}},
\end{equation}
where $\Gamma(x)$ is the gamma function.

\begin{figure}
  \resizebox{\hsize}{!}{\includegraphics[angle=0]{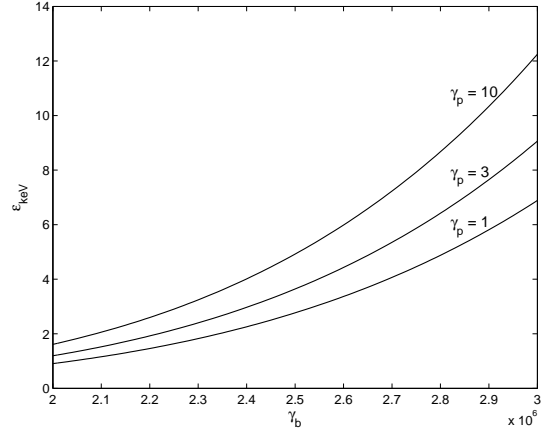}}
  \caption{Behaviour of energy of radiated
synchrotron photons with the beam electron Lorentz factors for
different values of plasma component Lorentz factors in case of
$H_{\perp}\gg F_{\perp}$. The set of parameters is $A_k = 1$, $M_8 =
1$, $A_{0.1} = 1$, $L = 10^{40}$erg/s, $\gamma_p\in[1;3;10]$, $n_b =
n_{_{GJ}}$.}\label{fig1}
\end{figure}

\section{Discussion}

As we see from the results of the previous section, the
Cherenkov-drift instability strongly influences the particle
distribution by means of the QLD and prevents the synchrotron
emission. In this section we apply the developed model to AGNs and
study the production of the nonthermal emission.

For this purpose we consider the light cylinder zone of the
magnetosphere. In this area the value of the magnetic induction
might be estimated in the framework of the equipartition
approximation. Thus, we assume that the magnetic field energy
density is of the order of plasma energy density, which defines the
magnetic induction
\begin{equation}
\label{b} B_{lc}\approx 5.5\times A_{0.1}\times L_{40}^{1/2}G.
\end{equation}
As we see the magnetic field is quite strong and since in this area
particles normally achieve very high Lorentz factors, $\sim
10^{6-7}$, the resulting synchrotron emission should be quite
efficient. In particular, relativistic electrons moving in a strong
magnetic field and having the pitch angles expressed by Eqs.
(\ref{pitch1},\ref{pitch2}), will emit the photons with energies
\citep{Lightman}
\begin{equation}\label{eps1}
\epsilon^{I}_{_{keV}}\approx 1.2\times 10^{-11}\gamma_b^2
B_{lc}\frac{1}{\sqrt{\pi}}
 \frac{p_{\perp_{0}}}{p_{\parallel}}.
\end{equation}
\begin{equation}\label{eps2}
\epsilon^{II}_{_{keV}}\approx 3\times 10^{-12}\gamma_b^2
B_{lc}\frac{\sqrt{\pi}}{\Gamma (\frac{5}{4})}
 \frac{p^{II}_{\perp_{0}}}{p_{\parallel}}.
\end{equation}

In the framework of the quasi-linear diffusion, the problem is
usually treated by means of the method of iteration. It is evident
that the instability has to be saturated by means of nonlinear
effects (the corresponding study is not in the scope of the paper)
and since there is some energy budget, it is clear that after the
process is relaxed, the energy of waves must be of the same order of
that of plasmas associated with the energy budget \citep{malmach}.
Therefore, in the framework of the paper we assume $W\sim
\gamma_bmc^2n_b$.

As a first example we consider the case when $H_{\perp}$ exceeds the
corresponding component of the radiation reaction force. By
combining Eqs.(\ref{incr},\ref{dkoef},\ref{pp01},\ref{eps1}) one can
study the emission characteristics. In Fig. \ref{fig1} we show the
dependence of energy of radiated synchrotron photons on the beam
electron Lorentz factors for different values of $\gamma_p$. The set
of parameters is $A_k = 1$, $M_8 = 1$, $A_{0.1} = 1$, $L =
10^{40}$erg/s, $\gamma_p\in[1;3;10]$, $n_b = n_{_{GJ}}$, where
$n_{_{GJ}}\equiv\Omega B_{lc}/(2\pi ec)$ is the Goldreich-Julian
number density of electrons \citep{rud}. From Eqs. (\ref{pp01}) one
obtains that the pitch angles vary in the range $\sim 3\times
10^{-3}-2\times 10^{-2}$rad, that is in a good agreement with the
assumption $H_{\perp}\gg F_{\perp}$ (see Eq. \ref{ratio}).

As it is evident from the plots the photon energy is a continuously
increasing function of $\gamma_b$, which is a natural consequence of
the fact that more energetic electrons emit more energetic photons.
$\epsilon_{keV}$ has the similar behaviour with respect to the
plasma component Lorentz factor. By increasing $\gamma_p$ the
corresponding photon energy increases as well. In particular, by
combining Eqs. (\ref{incr},\ref{om},\ref{dkoef},\ref{pp01}) one can
see that $\epsilon_{keV}\propto D_{\perp,\perp}^{1/2}\propto
\gamma_p^{1/4}$. As it is clear from the figure, the QLD guarantees
the emission in the keV energy domain ($X$-rays), which in turn must
be strongly correlated with the lower energy emission provided by
Cherenkov-drift instability. From Eqs. (\ref{ud},\ref{om}) one can
obtain energy of photons corresponding to the Cherenkov drift waves
\begin{equation}
\label{lowen} \epsilon_{eV}^{_{Ch}}\approx
0.06A_{0.1}\gamma_p^{-3/2}\times\left(\frac{n_b}{n_{_{GJ}}}\right)^{1/2}
\times\left(\frac{L}{10^{40}erg/s}\right)^{1/2}.
\end{equation}

\begin{figure}
  \resizebox{\hsize}{!}{\includegraphics[angle=0]{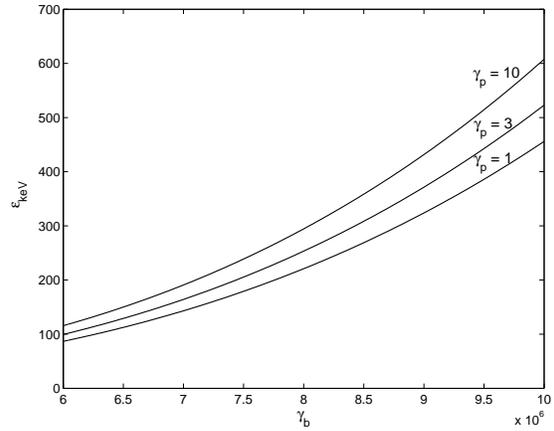}}
  \caption{Dependence of energy of synchrotron
photons on $\gamma_b$ for different values of plasma component
Lorentz factors in case of $H_{\perp}\ll F_{\perp}$. The set of
parameters is $A_k = 1$, $M_8 = 1$, $A_{0.1} = 1$, $L =
10^{40}$erg/s, $\gamma_p\in[1;3;10]$, $n_b =
n_{_{GJ}}$.}\label{fig2}
\end{figure}

\begin{figure}
  \resizebox{\hsize}{!}{\includegraphics[angle=0]{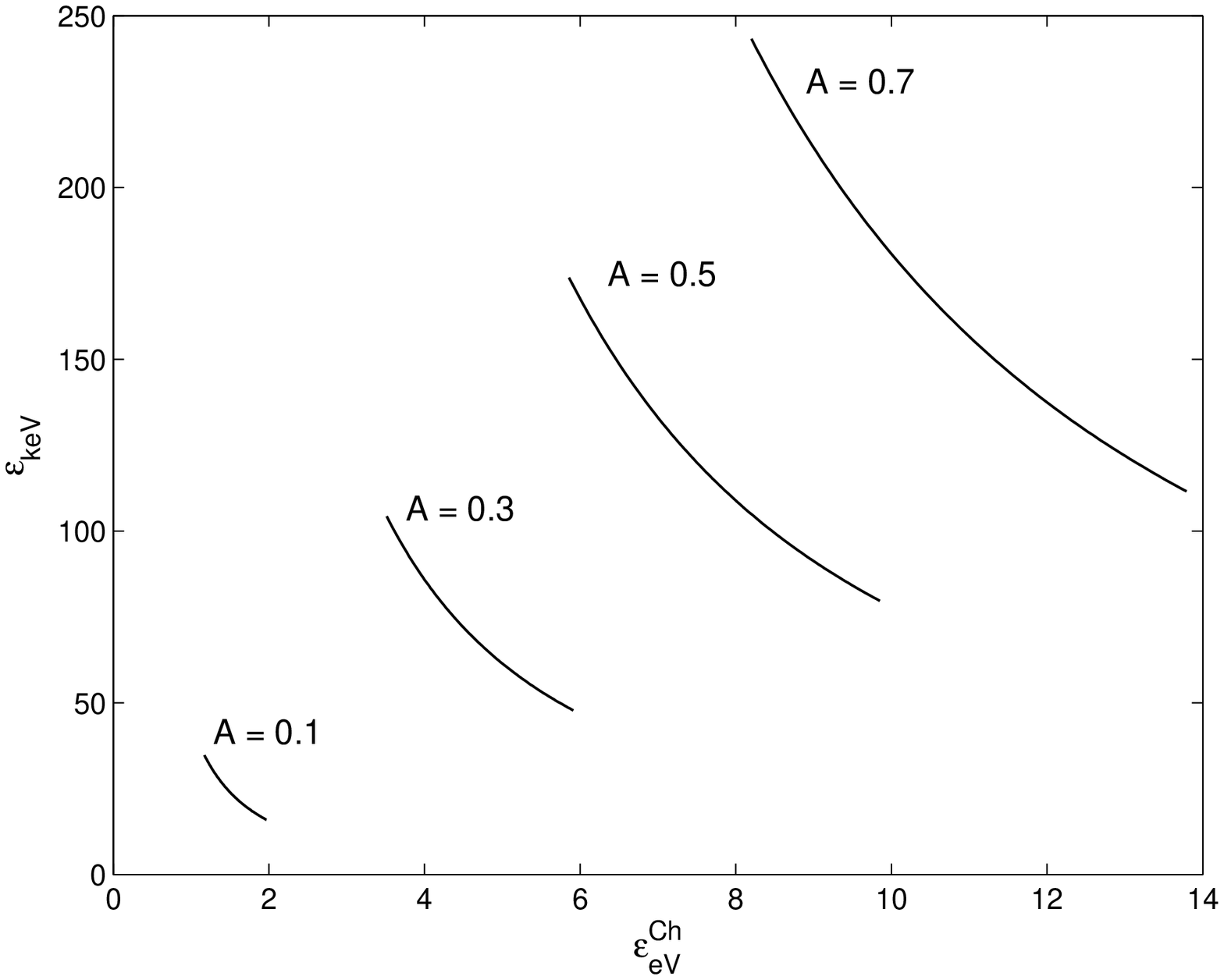}}
  \caption{Behaviour of $\epsilon_{keV}$ with $\epsilon^{Ch}_{eV}$ for different
values of $A$ in case of $H_{\perp}\gg F_{\perp}$. The bolometric
luminosity varies in the interval $[0.5-1]\times 10^{42}$erg/s. The
set of parameters is $A_k = 1$, $M_8 = 1$, $A_{0.1}\in[1;5;10]$,
$\gamma_b = 10^7$, $\gamma_p =1$, $n_b=n_{_{GJ}}$.}\label{fig3}
\end{figure}

We see that for the mentioned physical parameters the Cherenkov
instability produces the IR photons with energies $\sim (2\times
10^{-3}-6\times 10^{-2})$eV, and since the QLD is achieved by means
of the feedback of these waves on particles, it is evident that the
IR emission must be strongly connected to the soft $X$-ray radiation
($1-12$)keV.

Generally speaking, the instability is supposed to be efficient if
the corresponding growth rate is high enough. From Eq. (\ref{incr})
we obtain
\begin{eqnarray}
\label{incr1} \Gamma_k\approx 6.5\times 10^{-3}\times
\gamma_p^{-1/2}\times\frac{\gamma_b}{10^{7}}\times\nonumber\\
\times \left(\frac{n_b}{n_{_{GJ}}}\right)^{1/2}
\times\left(\frac{L}{10^{40}erg/s}\right)^{1/2}\; s^{-1}.
\end{eqnarray}
It is clear that the instability timescale, $t_{inst}\sim
1/\Gamma_k$ is of the order of $500$s. On the other hand, particles
are moving inside the magnetosphere and the kinematic timescale
(escape timescale) $t_{kin}\sim R_{lc}/c\approx 5\times 10^{3}$s
exceeds the instability timescale, which indicates high efficiency
of the Cherenkov-drift mechanism.

From Eq. (\ref{ratio}) it is clear that $H_{\perp}/F_{\perp}$ is
sensitive with the Lorentz factor of beam components and the pitch
angles (note that $\psi$ itself nontrivially depends on $\gamma_b$
see Eqs. (\ref{pitch1},\ref{pitch2})) and hence, for other
parameters the relation between forces may be different. Therefore,
it is reasonable to consider another limit, ($H_{\perp}\ll
F_{\perp}$) and see what happens in this particular situation.

In Fig. \ref{fig2} we show the dependence of energy of synchrotron
emission on beam Lorentz factors for different values of $\gamma_p$.
The set of parameters is $A_k = 1$, $M_8 = 1$, $A_{0.1} = 1$, $L =
10^{40}$erg/s, $\gamma_p\in[1;3;10]$, $n_b = n_{_{GJ}}$. The
difference from the previous case is $\gamma_b$, which varies from
$6\times 10^6$ to $10^7$. From Eq. (\ref{ratio}) it is clear that
for the mentioned parameters the following approximation,
$H_{\perp}\ll F_{\perp}$ is satisfied. From the plots we see that
$\epsilon_{keV}$ lies in the interval $\sim (100-600)$keV, which is
much higher than the photon energies shown in Fig. \ref{fig1} being
a result of a more steep dependance
$\epsilon_{keV}\propto\gamma_b^{13/4}$ (see
Eqs.(\ref{incr},\ref{dkoef},\ref{pp02},\ref{eps2})). Since the
expression of $\epsilon^{Ch}_{eV}$ does not depend on the beam
Lorentz factors, the hard $X$-ray radiation is strongly connected
with the corresponding Cherenkov-drift emission generated in the
same energy domain as in the previous case: IR band.

The emission pattern strongly depends on the values of the AGN
luminosity. In particular, in the framework of the model, we use the
equipartition magnetic field and the induction becomes dependent on
$L$. Such a dependance is motivated by the fact that the diffusion
coefficient behaves as, $1/{B}^8$ (see Eq. (\ref{dkoef})) which
leads to $\epsilon_{keV}\propto L^{-9/8}$ (see Eqs.
(\ref{pitch1},\ref{pitch2},\ref{eps1},\ref{eps2})). On the other
hand, unlike the results shown on the previous two graphs, in this
case $\epsilon^{Ch}_{eV}$ also depend on the AGN luminosity, and
since synchrotron and Cherenkov-drift emission are generated
simultaneously, it is interesting to investigate them both.

In Fig. \ref{fig3} we present the behaviour of the synchrotron
photon energy on the Cherenkov-drift emission for different values
of angular momentum. Unlike the previous cases it is worthwhile to
consider higher luminosity values: $[0.5-1]\times 10^{42}$erg/s. We
do not examine extreme luminous sources, because in this case the
Lorentz factors of relativistic electrons might be less than
considered in the present paper. The set of parameters is $A_k = 1$,
$M_8 = 1$, $A_{0.1}\in[1;5;10]$, $\gamma_b = 10^7$, $\gamma_p =1$,
$n_b=n_{_{GJ}}$. It is straightforward to show that the synchrotron
radiation reaction force is less than the force responsible for
conservation of the adiabatic invariant. It is clear from the plots
that $\epsilon_{keV}\left(\epsilon^{Ch}_{eV}\right)$ is a
continuously decreasing function, which is a direct result of the
fact that with bolometric luminosity $\epsilon_{keV}$ decreases
($\propto L^{-9/8}$), whereas the energies of Cherenkov-drift
photons behave as $L^{1/2}$ (see Eq. (\ref{lowen})). We also show
the plots for different values of angular momentum of the
supermassive black hole. As it is clear,  bigger the angular
momentum, the bigger the resulting emission energy for both
mechanisms. In particular, from Eqs.
(\ref{lowen},\ref{dkoef},\ref{pp02},\ref{eps2})) one can see that
$\epsilon_{keV}$ and $\epsilon^{Ch}_{eV}$ both are increasing
functions of $A$. According to the present results, we see that for
the mentioned parameters, the quasi-linear diffusion provides the
generation of synchrotron emission from the soft up to hard
$X$-rays: $\sim (10-250)$keV. These energies are strongly connected
with the Cherenkov-drift emission in the energy interval $\sim
(1-14)$eV, corresponding to emission from IR up to extreme UV.

\section{Summary}

The main aspects of the present work can be summarized as follows:
\begin{enumerate}

      \item In the present paper we studied the role of Cherenkov
      drift emission in maintaining the synchrotron regime despite
      the efficient energy losses.

      \item It is shown that in the magnetospheres of supermassive
      black holes the excited Cherenkov drift instability is
      efficient enough to influence the particle distribution by
      means of the quasi-linear diffusion.

      \item In the framework of the model, we examine equation
      governing the process of the QLD. It is shown that for physically
      reasonable parameters the synchrotron radiation reaction force prevails over other
      dissipative factors. By taking into account this fact the corresponding
      kinetic equation is analytically solved and the emission
      characteristics are estimated

      \item The emission pattern is investigated for a variety of
      physical parameters. We found that for typical AGNs the QLD might guarantee
      the excitation of strongly coupled Cherenkov drift emission in the eV domain
      and the synchrotron radiation in the keV energy band.

      \end{enumerate}

As we have already seen, the Cherenkov-drift instability might
guarantee the maintenance of the synchrotron emission process for a
variety of AGN. Another important issue we would like to address is
the problem of radiation spectral index, which, in some sense, will
complete the problem. It is necessary to investigate this task as
well, therefore, sooner or later we are going to examine it.

\section*{Acknowledgments}

We are very grateful to Prof. George Machabeli and Dr. David
Shapakidze for valuable discussions and helpful suggestions on the
manuscript.

\end{document}